\documentclass{kluwer}    

\usepackage{graphicx}
\usepackage{marvosym}

\newdisplay{guess}{Conjecture}

\begin{document}                                                                                   

\begin{article}

\begin{opening}         

\title{AA\,Dor --- An Eclipsing sdOB -- Brown Dwarf Binary} 
\author{Thomas \surname{Rauch}\email{Thomas.Rauch@sternwarte.uni-erlangen.de}}  
\runningauthor{Thomas Rauch}
\runningtitle{AA Dor --- An Eclipsing sdOB -- Brown Dwarf Binary}
\institute{Dr.-Remeis-Sternwarte, Sternwartstra\ss e 7, D-96049 Bamberg, Germany\\
           Institut f\"ur Astronomie und Astrophysik, Sand 1, D-72076 T\"ubingen, Germany}
\date{~}

\begin{abstract}
AA Dor is an eclipsing, close, post common-envelope binary (PCEB) consisting of a sdOB primary star 
and an unseen secondary with an extraordinary small mass ($M_2 \approx 0.066\,\mathrm{M_\odot}$) -- formally a
brown dwarf. The brown dwarf may have been a former planet which survived a 
common envelope (CE) phase and has even gained mass.

A recent determination of the components' masses from results of NLTE spectral analysis
and subsequent comparison to evolutionary tracks shows a discrepancy to
masses derived from radial-velocity and the eclipse curves. 
Phase-resolved high-resolution and high-SN spectroscopy was carried out in order to investigate
on this problem. 

We present results of a NLTE spectral analysis of the primary, an analysis of its orbital parameters,
and discuss possible evolutionary scenarios.
\end{abstract}

\end{opening}           
 
\section{Introduction}  

AA\,Dor (LB\,3459, $\alpha_{2000} = 05^\mathrm{h} 31^{\mathrm m} 40.349^\mathrm{s}$,
                   $\delta_{2000} = -69^\mathrm{o} 53' 02.18''$) is an eclipsing,
single-lined binary system (Kilkenny et al\@. 1978). 
It is a relatively bright ($m_{\mathrm V} = 11.138$), extremely blue foreground object ($d \approx 400\,\mathrm{pc}$,
Rauch 2000) of the Large Magellanic Cloud (Feast et al\@. 1960).
Kilkenny et al\@. (1979) measured a period of $P = 0.2615\,\mathrm{d}$ and found an inclination of
$i = 90.00^\mathrm{o}\pm 0.02^\mathrm{o}$. The duration of the primary eclipse is $\approx 22\,\mathrm{min}$ with 
$\Delta m_{\mathrm V} \approx 0.4\,\mathrm{mag}$. 
There is a reflection effect (cf\@. Hilditch et al\@. 1996) with $\Delta m_{\mathrm V} \approx -0.05\,\mathrm{mag}$
and a secondary eclipse with $\Delta m_{\mathrm V} \approx 0.06\,\mathrm{mag}$.

Photometric investigations and light-curve analyses (Kilkenny et al\@. 1978, 1979, 1981) as well as
evolutionary models (Paczynski 1980) have shown that AA\,Dor consists of a sdOB primary star 
and an unseen, nearly degenerate, hydrogen-rich dwarf star ($M_2 \approx 0.054\,\mathrm{M_\odot}$) of low temperature
($T_\mathrm{eff} \approx 3\,\mathrm{kK}$).
One hemisphere of the secondary is heated by the primary (up to 15 - 20\,kK) and responsible for the reflection
effect.

Regarding the evolutionary times of both components of AA\,Dor,  
Paczynski (1980) considered the primary to have only a hydrogen-burning shell and $M_1 \approx 0.36\,\mathrm{M_\odot}$ 
to be more likely than to have double-shell burning and $M_1 \approx 0.54\,\mathrm{M_\odot}$.

\section{Spectral Analyses of the Primary}

In order to derive further constraints for the system, Kudritzki et al\@. (1982) presented the first spectral
analysis of the primary by means of NLTE model atmosphere techniques, based on UV and optical spectra.
From their results (Tables~\ref{php}, \ref{mrt}), they concluded that
the primary appears to be a sdOB star with a hydrogen-burning shell and a degenerate helium core, 
whose surface composition is dominated by diffusion (which is responsible for the low helium abundance). 
The secondary is close to a degenerate configuration of solar composition.

\begin{table}[ht]
\caption[]{Photospheric parameters of the primary of AA\,Dor.
          }
\label{php}
\begin{tabular}{llr@{.}lr@{.}l}                                        
\hline  
authors &
\multicolumn{1}{c}{$T_\mathrm{eff}$ / kK} &
\multicolumn{2}{c}{$\log g$ (cgs)} &
\multicolumn{2}{c}{He/H (by number)} \\
\hline
Kudritzki et al\@. 1982 & $40\,^{+3}_{-2}$ & 5 & $30\pm 0.3$ & 0 & $003\,^{+0.002}_{-0.001}$ \\
Rauch 2000              & $42\pm 1$        & 5 & $21\pm 0.1$ & 0 & $0008\pm 0.1\,\mathrm{dex}$  \\
\hline
\end{tabular}
\end{table}

Rauch (1987) used AA\,Dor as an example to check implications of more detailed model atoms
and numerical approximations in the NLTE code of Werner (1986). In comparison to
Kudritzki et al\@. (1982), the line-profile fits to the hydrogen Balmer lines were significantly
improved due to the consideration of Stark line broadening for the bound-bound transitions of H 
in the calculation of the statistical equilibrium in more detail. However, for an unknown
reason at that time, these fits were still not perfect. The reason for this became clear
when Werner (1996) found that the neglect of metal opacities in the atmosphere calculation
results in the so-called Balmer-line problem (Napiwotzki \& Rauch 1994).

\begin{figure}[ht]
\centerline{\includegraphics[width=\textwidth]{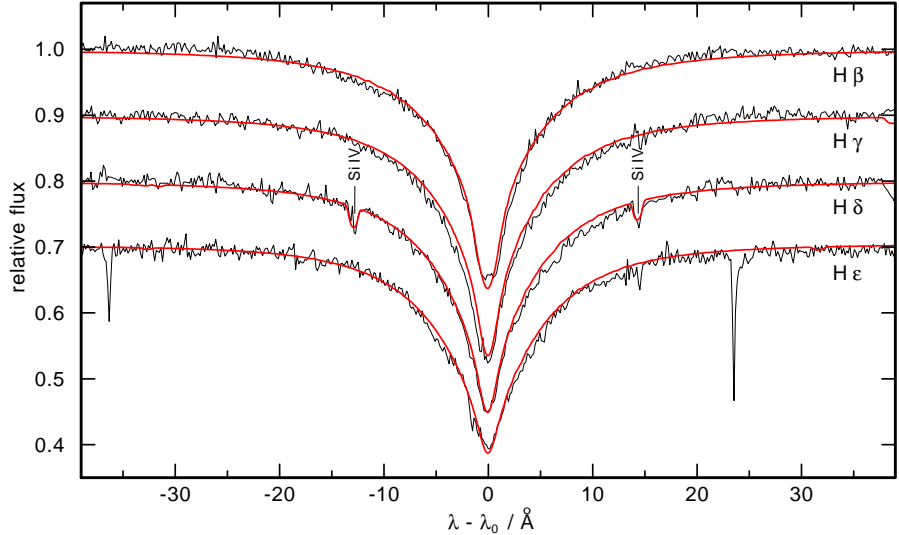}}
\caption{Theoretical line profiles of H\,$\beta$ -- H\,$\epsilon$ calculated from the
final H+He+C+N+O+Mg+Si+Fe+Ni model with $T_\mathrm{eff} = 42\,\mathrm{kK}$, $\log g = 5.21$,
and He/H = 0.0008 (by number) of Rauch (2000) compared with an ESO CASPEC spectrum of AA\,Dor.
(Other element abundances see Figure~\ref{diff}.) Note the signature of the reflection from the
secondary which is visible by the filled-in line core of H\,$\beta$.}
\label{balmer}
\end{figure}

\begin{figure}[ht]
\centerline{\includegraphics[width=\textwidth]{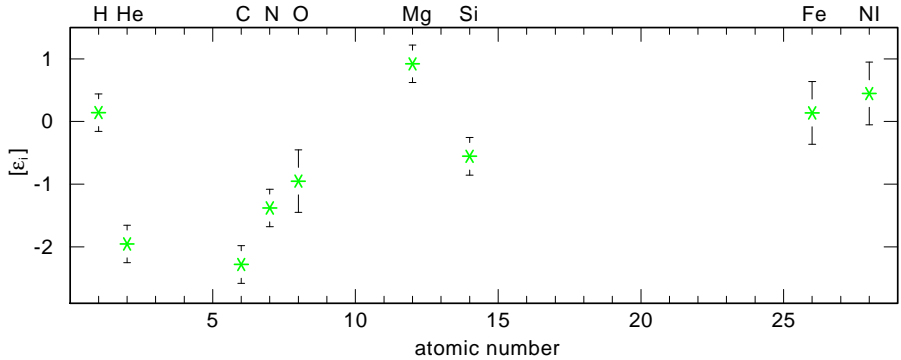}}
\caption{Photospheric abundances of the primary of AA\,Dor.
         $\left[\epsilon_\mathrm{i}\right]$ denotes $\log \left(\epsilon_\mathrm{i} / \epsilon_\odot\right)$, 
         with $\epsilon_\mathrm{i}$
         normalized to $\log \sum_\mathrm{i} \mu_\mathrm{i} \epsilon_\mathrm{i} = 12.15$
         (cf\@. Holweger 1979).
         Diffusion seems to be efficient and causes e.g\@. the strong depletion of helium.}
\label{diff}
\end{figure}

To make progress, Rauch (2000) presented then a detailed spectral analysis based on NLTE
model atmospheres which consider opacities of H, He, C, N, O, Mg, Si, Fe, and Ni with
326 individual atomic levels treated in NLTE, and 952 individual line transitions. 
About six million Fe and Ni lines are included in the calculations using a statistical approach
(Deetjen et al\@. 1999).
Based on high-resolution optical and ultraviolet spectra, the simultaneous evaluation of the ionization
equilibria of He\,{\sc i}/He\,{\sc ii}, C\,{\sc iii}/C\,{\sc iv}, N\,{\sc iii}/N\,{\sc iv}/N\,{\sc v},
and O\,{\sc iv}/O\,{\sc v} yields $T_\mathrm{eff} = 42\pm 1\,\mathrm{kK}$. Since the ionization
balances are very sensitive indicators for $T_\mathrm{eff}$, the error range is small. 
The results of this analysis are summarized in Table~\ref{php} and Figures\,\ref{balmer}, \ref{diff}.

\begin{table}[ht]
\caption[]{Masses and radii of the components of AA\,Dor.
           Masses are given in M$_\odot$, radii in R$_\odot$. 
           Results from spectral analyses: 
           K1982\,= Kudritzki et al\@. 1982, 
           R2000\,= Rauch 2000, 
           from light-curve and radial-velocity curve analyses: 
           P1980\,= Paczynski 1980,
           H1996\,= Hilditch et al\@. 1996 (on the assumption $M_1 = 0.5 \mathrm{M}_\odot$),
           K2000\,= Kilkenny et al\@. 2000 (on the assumption $M_1 = 0.3 \mathrm{M}_\odot$),
           H2003\,= Hilditch et al\@. 2003.
          }
\label{mrt}
\begin{tabular}{lr@{.}lr@{.}lr@{.}lr@{.}l}                                        
\hline
 &  
\multicolumn{4}{c}{primary} &
\multicolumn{4}{c}{secondary} \vspace{-2mm}\\
authors \vspace{-2mm}\\
\lcline{2-5}
\rcline{6-9}
\noalign{\smallskip}&  
\multicolumn{2}{c}{$M_1$} &
\multicolumn{2}{c}{$R_1$} &
\multicolumn{2}{c}{$M_2$} &
\multicolumn{2}{c}{$R_2$} \\
\hline
P1980 & 0 & 36             & 0 & 18                      & 0 & 054 & 0 & 10               \\
K1982 & 0 & 3              & 0 & 18                      & 0 & 04  & 0 & 09               \\
H1996 & 0 & 5              & \multicolumn{2}{c}{}        & 0 & 086 & \multicolumn{2}{c}{} \\
K2000 & 0 & 3              & \multicolumn{2}{c}{}        & 0 & 04  & \multicolumn{2}{c}{} \\
R2000 & 
\multicolumn{2}{c}{$0.324 - 0.336$} &
\multicolumn{2}{c}{$0.209 - 0.267$} &
\multicolumn{2}{c}{$0.065 - 0.067$} &
\multicolumn{2}{c}{$0.085 - 0.109$} \\
H2003 &  
\multicolumn{2}{c}{$0.33~ - 0.47~$} &
\multicolumn{2}{c}{$0.179 - 0.200$} &
\multicolumn{2}{c}{$0.064 - 0.082$} &
\multicolumn{2}{c}{$0.097 - 0.108$} \\ 
\hline
\end{tabular}
\end{table}

\section{Spectral analysis vs\@. light-curve analysis} 
\label{prob}

\begin{figure}[ht]
\centerline{\includegraphics[width=\textwidth]{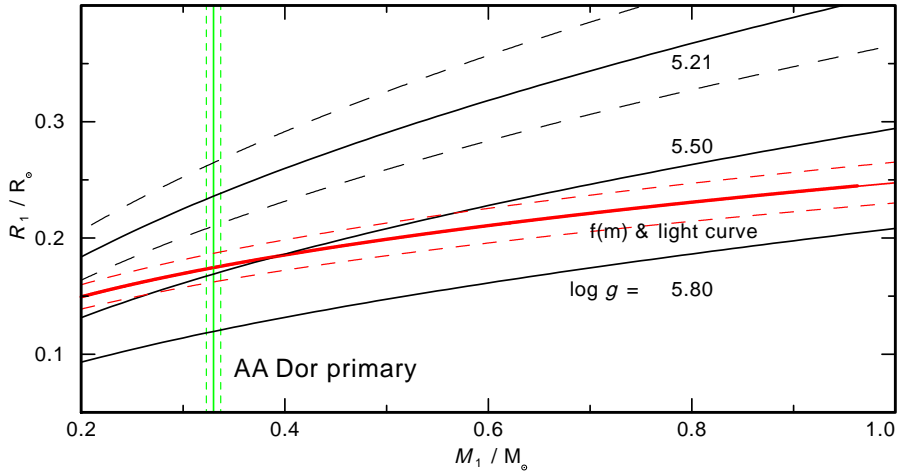}}
\caption{Mass-radius relation for the primary of AA\,Dor. Obviously, the
             solution from f(m) and light curve does not intersect with the result ($\log g = 5.21$) of Rauch (2000)
             and the mass value ($M_1 \approx 0.330\,\mathrm{M_\odot}$) determined from comparison to evolutionary models. 
             The dashed lines indicate the error ranges.}
\label{mrf}
\end{figure}

Although Rauch (2000) used advanced model atmospheres for the spectral analysis of the primary,
a ``$g$ problem'' appeared --- there is no realistic agreement in the mass-radius relation 
between his results and the solution of a mass function $f(m)$ and light-curve analysis (Figure~\ref{mrf}) ---
an intersection is found only at $M_1 < 0.2\,\mathrm{M_\odot}$ (within error limits) which seems to
be too low for a sdOB star.

The reason for this disagreement is unclear. Possible reasons may be too optimistic error ranges in
Rauch (2000) or in the analysis of light curve and radial-velocity curve, or that the theoretical
evolutionary models of Driebe et al\@. (1998) are not appropriate in the case of AA\,Dor since these
are post-RGB models for non-CE stars.

\begin{figure}[ht]
\centerline{\includegraphics[width=\textwidth]{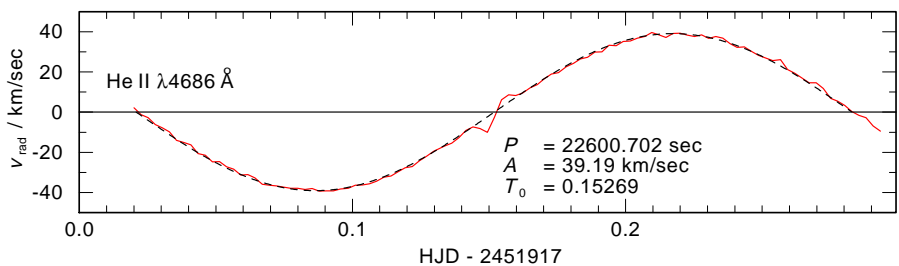}}
\caption{Radial-velocity curve of AA\,Dor. As an example, He\,{\sc ii} $\lambda$\,4686\,\AA\ is shown
here. Note the velocity jumps close to $T_0$ which are the result of the transit of the cool companion 
(Rossiter Effect, Rossiter 1924). The dashed line is a sine curve matching best the observation
with parameters derived by {\tt TRIPP} (Schuh et al\@. 2003).
}
\label{rv}
\end{figure}

The spectral analysis of Rauch (2000) was hampered by the long
exposure times (1 - 3h) of the available optical and ultraviolet spectra because smearing due to orbital
motion could not be separated properly from the rotational broadening. Thus, Rauch \& Werner (2003)
performed phase-dependent spectroscopy at ESO's Very Large Telescope and UVES
(UV-Visual Echelle Spectrograph) attached. 
One complete period of AA\,Dor was covered by 180\,sec exposures in order to mini\-mize the smearing effects
and to measure the radial-velocity curve (Figure~\ref{rv}). We employed {\tt TRIPP} (Schuh et al\@. 2003),
an IDL based aperture-photometry package for the reduction of CCD time series, to derive
the radial-velocity amplitude $A_1$ of the primary and the orbital period.
$A_1 = 39.19\pm 0.05\,\mathrm{km/sec}$ could be determined precisely. 
Although additional radial-velocity measurements of Hilditch et al\@. (1996) from January 1994 have been included in the
analysis by {\tt TRIPP} (together covering 9695 periods),
the precision of the period determination of ($P = 22600.702\,\mathrm{sec}$) can,
of course, not compete with results of Kilkenny et al\@. (2000, $P = 22597.03189\,\mathrm{sec}$).
Depending on the data distribution, i.e\@. many data points in 2001 and the large gap towards the few 1994 data, 
it appears that the result of {\tt TRIPP} simply ``misses'' one period.
It is worthwhile to note, that Kilkenny et al\@. (2000) evaluated more than
30\,000 eclipses in 26 years and were even able to set an upper limit of $dP/dn < 10^{-13} \mathrm{d/orbit}$ to the
period change.

From detailed line profile fits of He\,{\sc ii} $\lambda$\,4686\,\AA, Rauch \& Werner (2003) determined
the rotational velocity $v_\mathrm{rot} = 47\pm 5\,\mathrm{km/sec}$ of the primary. 
At {\tt http://astro.uni-tuebingen.de/\raisebox{1mm}{{\small $\sim$}}rauch/aador.html},
an illustrative
animation of the orbital motion and the phase-dependent spectral variation is available.
Since the circularization
und synchronization times for AA\,Dor are only of the order of hundreds of years and the CE
was lost some $5\cdot 10^5\,\mathrm{years}$ ago (Section~\ref{evo}), we can calculate $R_1 = 0.217 - 0.269\,\mathrm{R_\odot}$
for bound rotation which is in agreement with the results of Rauch (2000, Table\,\ref{mrt}).
Unfortunately, the main aim of this investigation was not reached. Due to the relatively poor quality of the
UVES spectra and problems in the data reduction of the \'echelle spectra, it was not possible to improve 
the determination of $g$.

Recently, Hilditch et al\@. (2003) presented new photometry data and an improved photometric model of AA\,Dor which
verified earlier results within smaller error ranges.

To summarize, the precise $g$ determination is still a crucial point for the understanding of the
evolution of AA\,Dor.

\section{Evolutionary status of AA\,Dor}
\label{evo}

Based on measurements of the light curve and the radial-velocity curve, Paczynski (1980) presented an early model with a 
primary of about $0.36\,\mathrm{M_\odot}$, $T_\mathrm{eff} = 64\,\mathrm{kK}$, with a hydrogen-burning shell and a 
degenerate helium core. The secondary of about $0.054\,\mathrm{M_\odot}$ is nearly degenerate, and
probably hydrogen rich. AA\,Dor had an initial period of $P \approx 3\,\mathrm{months}$ and a separation of 
$a > 100\,\mathrm{R_\odot}$. It experienced a CE phase, during which the separation was reduced to
$a \approx 1.3\,\mathrm{R_\odot}$. About $5\cdot 10^5\,\mathrm{years}$ ago, the CE was lost.
Since then, the primary is burning hydrogen and has shrunk from Roche lobe (RL) filling dimensions to its
present size, while the secondary is contracting on a gravitational contraction time scale. In less than
$10^6\,\mathrm{years}$, nuclear burning will cease, and both components will cool off.
In about $5\cdot 10^{10}\,\mathrm{years}$, $P$ will decrease to about 38 minutes before the secondary
fills its RL then and mass transfer onto the primary will result in a short-period cataclysmic variable.

Hilditch et al\@. (1996) assumed $M_1 = 0.5\,\mathrm{M_\odot}$ and calculated $M_2 = 0.086\,\mathrm{M_\odot}$.
The secondary has then $T_\mathrm{eff} \approx 2\,\mathrm{kK}$ and $\log g = 5.32$ in excellent agreement
with the lowest mass ZAMS models of Dorman et al\@. (1989). The primary had an initial mass of $4\,\mathrm{M_\odot}$
and has undergone ``late case B mass transfer'' (Iben \& Livio 1993). The lower spectroscopic mass of 
$M_1 = 0.330\,\mathrm{M_\odot}$ (Rauch 2000, Table~\ref{mrt}) suggests that AA\,Dor is a
``low mass case B'' (RL filled just before He ignition) system (Iben \& Livio 1993). 

With its extraordinary low mass of 
$M_2 = 0.066\,\mathrm{M_\odot} \approx 70$\,M\raisebox{-1mm}{{\small \Jupiter}} (Rauch 2000), 
the secondary lies formally within the brown-dwarf mass range ($0.013 - 0.08\,\mathrm{M_\odot}$) 
and is burning deuterium in its core but
it is even possible that this is a former planet ($M_2 < 0.05\,\mathrm{M_\odot}$) which has
survived the CE phase ($M_2 > M_\mathrm{crit} \approx 0.02\,\mathrm{M_\odot}$) and has gained mass via RL overflow (RLOF). 
An alternate scenario is that a planet over the same critical mass limit gains mass directly
from the wind and the envelope of the primary (Livio \& Soker 1984). 

However, due to the low mass of the system, all these scenarios have a severe problem --
loss of orbital energy and angular momentum, i.e\@. when the secondary once started to
spiral-in during the CE phase, there might be no way to avoid its collision with the
core of the primary. In other words, since the initial mass of the primary
was $> 1\,\mathrm{M_\odot}$ (just to arrive at its present evolutionary phase), then
an envelope of more than $0.7\,\mathrm{M_\odot}$ has to be entirely ejected by the
secondary of only $M_2 = 0.066\,\mathrm{M_\odot}$ which appears impossible
(Eggleton \& Kiseleva-Eggleton 2002).

Recently, Eggleton \& Kiseleva-Eggleton (2002) proposed a scenario which appears not generally to end up with a
merger described above. In a first stage, a ``case AM'' (M: mass-loss dominated) mass transfer
takes place. The primary may either loose 75\% of its mass without ever filling its RL
while evolving to a red subgiant and then to a hot subdwarf
or two minor episodes of RLOF occur with a substantial detached phase. Later,
``case AL'' (L: late overtaking) mass transfer follows, where the secondary fills its RL, and
initiates reverse mass transfer. In the following CE phase, there will be a spiral-in which,
depending on the remaining envelope mass, ends with a merger or a detached close binary.
In the case of AA\,Dor, the binary would start with 
$M_1 \approx 1.0\,\mathrm{M_\odot}$, $M_2 \approx 0.05\,\mathrm{M_\odot}$, and $P \approx 20\,\mathrm{d}$. 
The secondary spins up the primary on its way up the giant branch and thus,
there will be a substantial mass loss combined with minimum angular-momentum loss. 
RLOF starts when the system arrives at 
$M_1 \approx 0.3\,\mathrm{M_\odot}$, $M_2 \approx 0.05\,\mathrm{M_\odot}$, and $P\,\approx\,60\,\mathrm{d}$. 
Since the envelope mass is now $\lsim\,0.05\,\mathrm{M_\odot}$, it may be expelled by the
secondary without spiraling in to the core of the primary.
\vspace{-1mm}

\section{Conclusions}

The evolutionary scenario of the PCEB AA\,Dor is still unclear, although Eggleton \& Kiseleva-Eggleton (2002)
have little doubt to be able to adjust their model (see Section~\ref{evo}) to obtain good agreement with AA\,Dor.
However, little is known about the secondary. Thus, it would be a valuable approach to hunt
for weak spectral signatures of the secondary (cf\@. Rucinski 2002) in high-resolution and
high-SN spectra or to try to measure its radial-velocity curve using the reflected light.

A disagreement remains between the results of spectral analysis and light curve and radial-velocity curve
analysis (see Section~\ref{prob}). The validity of the latter has been verified recently
by Hilditch et al\@. (2003) and thus, it appears likely that the spectral analysis yields ---
by an unknown reason so far --- a too-low value of $g$. Since the decrement of the hydrogen
Balmer series is very sensitive to variation of $g$ (e.g\@. Rauch et al\@. 1998), the analysis
of medium-resolution and very high-SN optical spectra, covering the wavelength range down to the Balmer edge,
would be a suitable way to attack this problem.  
\vspace{-2mm}

\begin{acknowledgements}
This research was supported by the DLR under grants 50\,OR\,9705\,5 and 50\,OR\,0201.
I like to thank Tony Lynas-Gray who sent me a preprint of the Hilditch et al\@. (2003) paper
just in time before this conference.
\end{acknowledgements}
\vspace{-2mm}


\theendnotes

\end{article}

\end{document}